\documentclass[aps,prl, twocolumn,superscriptaddress,amsmath,amssymb,floatfix]{revtex4-2}

\usepackage{graphicx}
\usepackage{dcolumn}
\usepackage{bm}
\usepackage{xcolor}
\usepackage{hyperref}
\hypersetup{
    colorlinks=true,
    linkcolor=blue,
    filecolor=magenta,
    urlcolor=blue,
    citecolor=blue,
}

\graphicspath{{figs/}}

\renewcommand{\Re}{\operatorname{Re}}
\renewcommand{\Im}{\operatorname{Im}}
\newcommand{\bra}[1]{\langle #1 |}

\newcommand{\braket}[2]{\langle #1 | #2 \rangle}
\newcommand{\orcid}[1]{\href{https://orcid.org/#1}{\includegraphics[width=7pt]{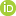}}}

\begin{document}

\title{Gouy Phase across PT-Symmetry Breaking in Non-Hermitian Dirac Systems}

\author{Dirlan J. C. de Morais Sousa~\orcid{0009-0007-7854-4445}}
\affiliation{Departamento de F\'isica, Universidade Federal do Piau\'i (UFPI), 64049-550 Teresina, PI, Brazil}

\author{Lucas S. Marinho~\orcid{0000-0002-2923-587X}}
\email{lucas.marinho@ufpi.edu.br}
\affiliation{Departamento de F\'isica, Universidade Federal do Piau\'i (UFPI), 64049-550 Teresina, PI, Brazil}

\author{Giandomenico Palumbo~\orcid{0000-0003-1303-1247}}
\email{giandomenico.palumbo@stp.dias.ie}
\affiliation{CFisUC, Department of Physics, University of Coimbra, Rua Larga, 3004-516 Coimbra, Portugal}

\author{Irismar G. da Paz~\orcid{0000-0002-9613-9642}}
\email{irismarpaz@ufpi.edu.br}
\affiliation{Departamento de F\'isica, Universidade Federal do Piau\'i (UFPI), 64049-550 Teresina, PI, Brazil}

\date{\today}

\begin{abstract}
We study relativistic beam-like wave packets governed by a quasi-Hermitian massive Dirac Hamiltonian and uncover anomalous Gouy-phase behavior in non-Hermitian dynamics. We show that the Gouy phase provides a sensitive probe of the global $\mathcal{PT}$-symmetry-breaking threshold: it remains purely real in the globally unbroken, quasi-Hermitian regime, while, after crossing the exceptional point, the Gouy phase changes sign and acquires an imaginary component. At the exceptional point, the Gouy-phase variation vanishes in the small-mass limit but becomes maximal for large masses, revealing a counterintuitive crossover from effectively classical to increasingly wave-like quantum behavior. We propose an experimental scheme to measure the components of the non-Hermitian Gouy phase in the broken regime by monitoring the attenuation of a light beam propagating through a lossy waveguide. These results highlight the potential of the non-Hermitian Gouy phase for photonic applications, including the determination of threshold conditions and the design and control of systems with gain and loss.
\end{abstract}

\maketitle

\textit{Introduction} - 
The Gouy phase is a fundamental wave phenomenon, originally identified as an additional phase shift acquired by a converging light wave upon propagation through its focal region \cite{Gouy1890_book, Gouy1890_CR}. Although commonly associated with Gaussian beams, it is a general consequence of transverse spatial confinement, arising from either focusing or diffraction \cite{Boyd1980, Feng2001, SimonMukunda1993}. Its accumulated value depends on the dimensionality and geometry of the confinement, reaching $\pi/2$ for cylindrical waves and $\pi$ for spherical waves upon propagation through focus, with corresponding fractional shifts for diffracted waves \cite{Boyd1980, Feng2001}. Beyond conventional optics, the Gouy phase has been observed in a broad variety of wave phenomena, including water and acoustic waves, as well as surface-plasmon and phonon-polariton excitations \cite{Chauvat2003, Holme2003, Zhu2007, Feurer2002}, highlighting its universal character. Its physical origin has been related to the geometry of focusing, geometric phases, and the uncertainty principle \cite{Boyd1980, Feng2001, SimonMukunda1993, Subbarao1995}, while its contribution to the resonance condition of optical cavities plays an important role in determining laser resonant frequencies \cite{Beijersbergen1993}. Its mode-dependent character has enabled applications in optical mode conversion, sorting, and manipulation \cite{Beijersbergen1993, Zhou2017, Gu2018}. More recently, the quantum Gouy phase, characterized by a photon-number-dependent phase evolution, has been experimentally demonstrated and explored for quantum-enhanced applications \cite{Kawase2008, Hiekkamaki2021, Hiekkamaki2022}.

In parallel, non-Hermitian quantum mechanics has matured from a formal curiosity into a central theme of contemporary physics. Bender and Boettcher's discovery that  ($\mathcal{PT}$)-symmetric Hamiltonians (namely systems invariant under parity ($\mathcal{P}$) and time-reversal ($\mathcal{T}$) symmetry) can possess entirely real spectra~\cite{BenderBoettcher1998,BenderBrodyJones2002}, together with systematic theory of pseudo-Hermiticity~\cite{Most2002a,Most2002b,Most2002c,Ashida2020,Zhu2021} and the earlier quasi-Hermitian framework~\cite{Scholtz1992}, established that a consistent unitary quantum theory can be built around a non-Hermitian Hamiltonian provided one adopts the correct, metric-deformed inner product. The defining feature of these systems is the \emph{exceptional point} (EP): a non-Hermitian degeneracy at which two or more eigenvalues \emph{and} their eigenvectors coalesce, leaving a defective Hamiltonian with a Jordan block~\cite{Heiss2012}. EPs and non-Hermitian phases underlie a now-vast theoretical and experimental phenomenology in photonics, acoustics, electronics and condensed matter physics---unidirectional invisibility, loss-induced transparency, $\mathcal{PT}$-symmetric lasing, exceptional-point-enhanced sensing, quench dynamics and non-Hermitian skin effect~\cite{ElGanainy2018,MiriAlu2019,Ozdemir2019,Ruter2010,Hodaei2017,Wiersig2014,Para2021, Song2019,Strasser2023,Rahul2026}.

Extending the Gouy phase study to non-Hermitian systems is therefore of particular interest, as these systems provide a natural framework for describing gain and loss in open quantum dynamics \cite{Ashida2020, ElGanainy2018}. While the Gouy phase can, in principle, be investigated in open quantum systems through the density matrix (or cross-spectral
density) formalism, such an approach is considerably more challenging because the global phase is eliminated in the construction of the density matrix. Consequently, extracting the Gouy phase requires an interferometric scheme capable of revealing the relative phase between different quantum states (or light beams) \cite{Pang2012, daPaz2016}. In contrast, the non-Hermitian Hamiltonian formalism provides direct access to exceptional points (EPs) and $\mathcal{PT}$-symmetry-breaking transitions, enabling a deeper understanding of the behavior of the Gouy phase in these regimes \cite{BenderBoettcher1998, Heiss2012}.

Our results show that the Gouy phase can serve as a sensitive indicator of $\mathcal{PT}$-symmetry breaking: although it remains purely real at exceptional points where $\mathcal{PT}$ symmetry is preserved, it acquires a nonzero imaginary component once the system enters the $\mathcal{PT}$-broken regime. Therefore, the emergence of a complex Gouy phase constitutes a clear signature of $\mathcal{PT}$-symmetry breaking. More fundamentally, this result extends the geometric character of the Gouy phase to the non-Hermitian domain, where it can be naturally understood within a biorthogonal geometric framework \cite{Brody2014}. While its real part retains the usual Gouy phase-shift interpretation, its imaginary part describes the non-unitary amplitude evolution associated with gain and loss. Thus, $\mathcal{PT}$-symmetry breaking manifests itself not only through spectral and amplitude changes, but also through a qualitative transformation of the geometry encoded in wave propagation. Also, given the central role of the Gouy phase in determining resonance conditions and laser threshold properties \cite{Beijersbergen1993}, the non-Hermitian Gouy phase can be essential to determine the threshold power, with potential implications for the design and control of photonic devices operating in the presence of gain and loss as studied in \cite{Kullig2025}. At the EP, the Gouy phase displays an equally striking mass-dependent behavior: its variation tends to vanish in the small-mass limit, whereas it approaches its maximal variation for large masses. This suggests an unusual crossover in the physical behavior of the system. For small masses, the suppression of the Gouy phase indicates an effectively classical behavior, while, counterintuitively, increasing the mass enhances the Gouy-phase signature, indicating an increasingly quantum behavior of the system. Thus, rather than being washed out in the large-mass regime, the Gouy-phase contribution to the wave dynamics becomes maximal.

\textit{Unbroken and Broken $\mathcal{PT}$-Symmetry: from real to complex Gouy Phase} - 
We study relativistic beam-like wavepackets governed by a non-Hermitian massive Dirac Hamiltonian in $(1+1)$ dimensions. The model is \emph{pseudo-Hermitian} and more precisely quasi-Hermitian, meaning there exists an invertible and positive-define Hermitian metric operator $\eta$ such that $H^\dagger = \eta\, H \, \eta^{-1}$, which allows a consistent probabilistic interpretation via the $\eta$-inner product $\langle \psi | \phi \rangle_\eta = \langle \psi | \eta | \phi \rangle$. We consider the Dirac Hamiltonian with a non-Hermitian deformation and natural units ($\hbar = c = 1$), i.e., 
\begin{equation}
    H = \alpha p + \beta m + i \mu \gamma_3, \qquad p=-i\partial_x, \quad \mu\in\mathbb{R},
    \label{eq:H_def_I}
\end{equation}
where $m$ is the mass and $\mu$ controls the non-Hermitian deformation. We choose the $2\times 2$ representation $\alpha = \sigma_x, \quad \beta = \sigma_z, \quad \gamma_3 = i \alpha \beta = \sigma_y$, so that
\begin{equation}
    H =
    \begin{pmatrix}
        m & p+\mu \\[1mm]
        p-\mu & -m
    \end{pmatrix},
    \label{eq:H_matrix}
\end{equation}
which is non-Hermitian for $\mu \neq 0$. We define parity and time reversal as
$\mathcal{P}=\sigma_z\mathsf{P}_x$ and
$\mathcal{T}=\sigma_z K$, respectively, where
$\mathsf{P}_x\psi(x,t)=\psi(-x,t)$ and $K$ denotes complex
conjugation. Thus, $\mathcal{PT}=\mathsf{P}_xK$ and
$(\mathcal{PT})H(\mathcal{PT})^{-1}=H$. While parity and time reversal
separately map $\mu\to-\mu$, their combined action leaves the
Hamiltonian invariant.

Diagonalizing \eqref{eq:H_matrix} gives the energy dispersion $E_\pm(p) = \pm \sqrt{p^2 + m^2 - \mu^2}$. The energies are real for all $p$ if $|\mu| < |m|$, which defines the \emph{unbroken quasi-Hermitian regime}. At $|\mu| = |m|$ the gap closes at $p=0$, and for $|\mu| > |m|$ the spectrum becomes complex. In the unbroken regime, it is convenient to map $H$ to a Hermitian Dirac Hamiltonian (Hermitian partner) $h = \alpha p + \beta M$, where $M = \sqrt{m^2 - \mu^2}$. Now, introducing the invertible operator $\rho = e^{\frac{\theta}{2}\alpha}, \ \tanh \theta = \mu/m$, we have the similarity transformation $H = \rho^{-1} h \, \rho$, $\eta = \rho^\dagger \rho = e^{\theta \alpha}$. This ensures pseudo-Hermiticity and allows all physical overlaps to be computed equivalently in the Hermitian partner theory. Notice that, although the Hamiltonian is $\mathcal{PT}$ symmetric for
all real values of $\mu$, it is globally quasi-Hermitian if and only if
its $\mathcal{PT}$ symmetry is unbroken throughout the spectrum,
namely for $|\mu|<|m|$. For $|\mu|>|m|$, the spectrum develops $\mathcal{PT}$-broken
momentum sectors, and a global positive-definite metric operator no
longer exists.
Now, we study the biorthogonal time evolution under the pseudo-Hermitian Dirac Hamiltonian $H$, i.e., $i \partial_t |\Psi_R(t)\rangle = H |\Psi_R(t)\rangle$ \cite{Brody2014}. In the unbroken quasi-Hermitian regime $|\mu| < |m|$, the positive-definite metric $\eta$ allows defining the associated left state $|\Psi_L(t)\rangle= \eta |\Psi_R(t)\rangle, \langle\Psi_L(t)| = \langle\Psi_R(t)| \eta$, so that physical overlaps are $\langle\Psi_R|\eta|\Psi_R\rangle$ and the $\eta$-norm is conserved:
\begin{equation}
    \frac{d}{dt} \langle \Psi_R(t)|\eta|\Psi_R(t)\rangle = i \langle \Psi_R(t)|(H^\dagger \eta - \eta H)|\Psi_R(t)\rangle = 0.
\end{equation}
We normalize states so that $\langle \Psi_R(t)|\eta|\Psi_R(t)\rangle = 1$.

The similarity transformation introduced above allows defining the Hermitian-picture state (the mapping) $|\Phi(t)\rangle= \rho |\Psi_R(t)\rangle$, which satisfies standard Hermitian evolution $i \partial_t |\Phi(t)\rangle = h |\Phi(t)\rangle$. All physical overlaps are preserved $\langle \Psi_R^{(a)}(t)|\eta|\Psi_R^{(b)}(t)\rangle = \langle \Phi^{(a)}(t)|\Phi^{(b)}(t)\rangle$. This mapping simplifies the construction of localized packets and computation of Gouy and geometric phases: one can work in the Hermitian partner theory, then map results back to the pseudo-Hermitian setting without ambiguity in normalization or interference phases. 
In the following, we construct localized Dirac wavepackets, study their phase evolution, and identify the Gouy phase in the pseudo-Hermitian setting. All results are formulated consistently using the $\eta$-inner product and the Hermitian mapping. We construct beam-like relativistic Gaussian wavepackets using the Hermitian partner Hamiltonian $h$ and map them to the pseudo-Hermitian theory via the similarity transformation. Consider a positive-energy wavepacket centered at momentum $p_0$ with energy $E_0 = \sqrt{p_0^2 + M^2}$. The group velocity and curvature at $p_0$ are $v_0 = p_0/E_0$ and $\beta_2= M^2/E_0^3$. Now, writing the Hermitian spinor in carrier-envelope form
\begin{equation}
    \Phi(x,t) \approx u_0 \, A(\xi,t) \, e^{i(p_0 x - E_0 t)}, \qquad \xi = x - v_0 t,
\end{equation}
with $u_0$ the positive-energy eigenspinor of $h(p_0)$ and $A(\xi,t)$ varying slowly, the envelope satisfies the paraxial like equation
\begin{equation}
    i \partial_t A(\xi,t) = - \frac{\beta_2}{2} \, \partial_\xi^2 A(\xi,t).
\end{equation}

Choosing an initial Gaussian envelope of width $w_0$,
\begin{equation}
    A(\xi,0) = \frac{1}{(\pi w_0^2)^{1/4}} \exp\left(-\frac{\xi^2}{2 w_0^2}\right),
\end{equation}
the solution is
\begin{equation}
    A(\xi,t) = \frac{1}{(\pi w_0^2)^{1/4}} \frac{1}{\sqrt{q(t)}} \exp\!\left[-\frac{\xi^2}{2 w_0^2 q(t)}\right],
\end{equation}
with
\begin{equation}
    q(t) = 1 + i \frac{t}{t_d}, \quad t_d = \frac{w_0^2}{\beta_2}.
\end{equation}

The real packet width and the Gouy phase associated with the wavepacket spreading are
\begin{equation}
    w(t) = w_0 \sqrt{1 + \left(\frac{t}{t_d}\right)^2}, \;\;\gamma_G(t) = \frac{1}{2} \arctan\left(\frac{t}{t_d}\right).
    \label{eq:width_and_gouy}
\end{equation}

The corresponding pseudo-Hermitian right and left states are
\begin{equation}
    |\Psi_R(t)\rangle = \rho^{-1} |\Phi(t)\rangle, \qquad |\Psi_L(t)\rangle = \eta |\Psi_R(t)\rangle,
\end{equation}
so that all probabilities and interference observables are computed with the $\eta$-inner product. The Gouy phase $\gamma_G(t)$ is preserved under this mapping. The physical consequences are that in a pseudo-Hermitian Dirac theory, the deformation modifies the inner product rather than breaking probability conservation. Experimentally relevant quantities are therefore expressed using the metric operator $\eta$ and the associated biorthogonal structure.

For a right state $\Psi_R(x,t)$ evolving under $H$,
\begin{equation}
    i \partial_t \Psi_R(x,t) = H \, \Psi_R(x,t),
\end{equation}
the physical (biorthogonal) probability density is $I(x,t) = \Psi_R^\dagger(x,t) \, \eta \, \Psi_R(x,t)$, and the conserved $\eta$-norm reads $\int_{-\infty}^{+\infty} dx \, I(x,t) = \langle \Psi_R(t)|\eta|\Psi_R(t)\rangle = 1$.

The non-Hermitian deformation acts, in the unbroken regime, as a continuous \emph{mass renormalization} $m\to M$: it does not spoil unitarity, it softens the Dirac gap. The reduction $m\to M$ already signals what happens at $\mu=m$: the effective mass vanishes. The degeneracy is, however, far more singular than a massless Dirac point. Setting $p=0$ and $\mu=m$ in Eq.~\eqref{eq:H_def_I}, the resulting Hamiltonian is not diagonalizable but is similar to a single $2\times2$ Jordan block with zero eigenvalue. The two eigenvectors coalesce into the single self-orthogonal vector defining a second-order EP~\cite{Heiss2012,MiriAlu2019}. This is the structural reason the geometric Gouy phase will behave anomalously: at the EP the state space itself degenerates. The naive norm $\braket{\Psi_R}{\Psi_R}$ is \emph{not} conserved: because $H$ is non-normal, the right eigenvectors are not orthogonal and the naive norm oscillates even though the spectrum is real. This is the relativistic analogue of the metric-dependent probability that underlies all consistent non-Hermitian quantum mechanics~\cite{Most2002a,Scholtz1992}: only the $\eta$-weighted current is physically meaningful.

From now, we analyze the broken regime $|\mu|>|m|$, where the energy spectrum is no longer purely real for all momenta. In this regime a positive-definite metric $\eta$ implementing $\eta$-unitarity generally does not exist globally, and the geometric phase defined from biorthogonal overlaps become \emph{complex}. We focus on the case of narrow-band wavepackets centered at a momentum $p_0$ in the region where the carrier energy remains real. From the dispersion relation $E_\pm(p)=\pm\sqrt{p^2+m^2-\mu^2}$, we define $\kappa=\sqrt{\mu^2-m^2}>0$, so that $E_\pm(p)=\pm\sqrt{p^2-\kappa^2}$. The carrier energy is real for $|p|>\kappa$, and purely imaginary for $|p|<\kappa$. We will construct a narrow-band packet with mean momentum $p_0$ satisfying $|p_0|>\kappa$ and bandwidth $\Delta p$ small enough that the support stays in the real-carrier region. At the threshold momenta $p=\pm\kappa$ one has $E_\pm=0$ and the eigenvalue coalescence is accompanied by a loss of diagonalizability (an exceptional point in momentum space). Therefore, we assume $p_0$ is not taken too close to $\pm\kappa$, so that a smooth second-order expansion around $p_0$ is valid.

Because the model is not quasi-Hermitian in the broken regime, we do not use a global positive metric $\eta$. Instead, we work with instantaneous right/left states. Let $|\Psi_R(t)\rangle$ evolve as
\begin{equation}
    i\partial_t|\Psi_R(t)\rangle = H|\Psi_R(t)\rangle,
    \label{eq:time_evolution_broken}
\end{equation}
and define the associated left state $|\Psi_L(t)\rangle$ as a solution of the adjoint evolution
\begin{equation}
    i\partial_t|\Psi_L(t)\rangle = H^\dagger|\Psi_L(t)\rangle,
    \label{eq:left_evolution_broken}
\end{equation}
with the biorthogonal overlap $\langle \Psi_L(t)|\Psi_R(t)\rangle$ fixed by a chosen normalization/gauge.

We now construct a narrow-band, positive-energy packet, but without relying on the quasi-Hermitian mapping. Take the positive-energy branch
\begin{equation}
    E(p)=+\sqrt{p^2-\kappa^2},\qquad |p|>\kappa,
    \label{eq:E_positive_broken}
\end{equation}
and define $E_0=E(p_0),\; v_0=p_0/E_0,\;\beta_2=-\kappa^2/E_0^{3}$. Note that for $|p_0|>\kappa$ one has $E_0\in\mathbb{R}$ and therefore $v_0,\beta_2\in\mathbb{R}$ as long as we stay away from the exceptional points.

Write the wavepacket in carrier-envelope form
\begin{equation}
    \Psi_R(x,t)\approx u_0\,A(\xi,t)\,e^{i(p_0x-E_0 t)}, \qquad \xi=x-v_0 t,
    \label{eq:carrier_envelope_broken}
\end{equation}
with $u_0$ a (right) positive-energy eigenspinor evaluated at $p_0$ and $A$ a slowly varying envelope. Expanding the dispersion to second order yields the same effective paraxial like envelope equation as in the Hermitian case, but the \emph{interpretation} differs because the full evolution is not unitary and the physically relevant interference must be read out using biorthogonal overlaps.

In the real-carrier window one has $\beta_2\in\mathbb{R}$ and hence $t_d\in\mathbb{R}$, so the \emph{envelope} evolution preserves the familiar real width law given by Eq.~\eqref{eq:width_and_gouy}. However, the \emph{Gouy phase readout} is now naturally defined by the biorthogonal connection, and in general it yields a complex quantity even when the carrier energy is real. Accordingly, we define the \emph{complex Gouy phase} associated with spreading as:
\begin{equation}
    \Gamma_G(t)=\Re \Gamma_G(t)+i\,\Im \Gamma_G(t).
    \label{eq:complex_gouy_def}
\end{equation}
For the Gaussian family, a convenient operational characterization is obtained by expressing overlaps between different times in terms of $q(t)=|q(t)|e^{i\phi(t)}$, with $|q(t)|=\sqrt{1+(t/t_d)^2}$ and $\phi(t)=\arctan(t/t_d)$. Using the connection collapses onto its logarithmic derivative, $\bra{\Psi_L}\partial_t\Psi_R\rangle/\braket{\Psi_L}{\Psi_R}=-\tfrac12\dot q(t)/q(t)$, the phase separates into two physically distinct pieces,
\begin{align}
    \Re\,\Gamma_G(t) &= \frac{1}{2}\arctan\!\left(\frac{t}{t_d}\right), \label{eq:reGamma} \\
    \Im\,\Gamma_G(t) &= -\frac{1}{4}\ln\!\left[1+\left(\frac{t}{t_d}\right)^2\right]. \label{eq:imGamma}  
\end{align}
The real part reproduces the Gouy phase of the unbroken regime but, because $t_d$ now changes sign with $M^2<0$, its accumulation \emph{inverts} relative to the unbroken regime, a directly observable phase reversal across the EP. The imaginary part is new and exact: it is a \emph{logarithmic attenuation} (or amplification) that has no Hermitian counterpart, encoding the gain/loss imbalance of the broken phase. The two are not independent; they are the real and imaginary shadows of the single analytic object, locked together by the Kramers-Kronig-like structure of $\ln q(t)$. Interesting Remarks: (i) The analysis above applies to packets whose spectral weight remains in the real-carrier region $|p|>\kappa$. If the packet overlaps the interval $|p|<\kappa$, then parts of the spectrum are imaginary and exponential growth/decay dominates, requiring a separate treatment; (ii) Close to the exceptional points $p=\pm\kappa$, the curvature $\beta_2$ becomes singular and the narrow-band expansion breaks down; a uniform treatment would require retaining higher orders or working directly with the exact evolution near the defective point.

\begin{figure}[!ht]
\centering
\includegraphics[width=1.0\linewidth]{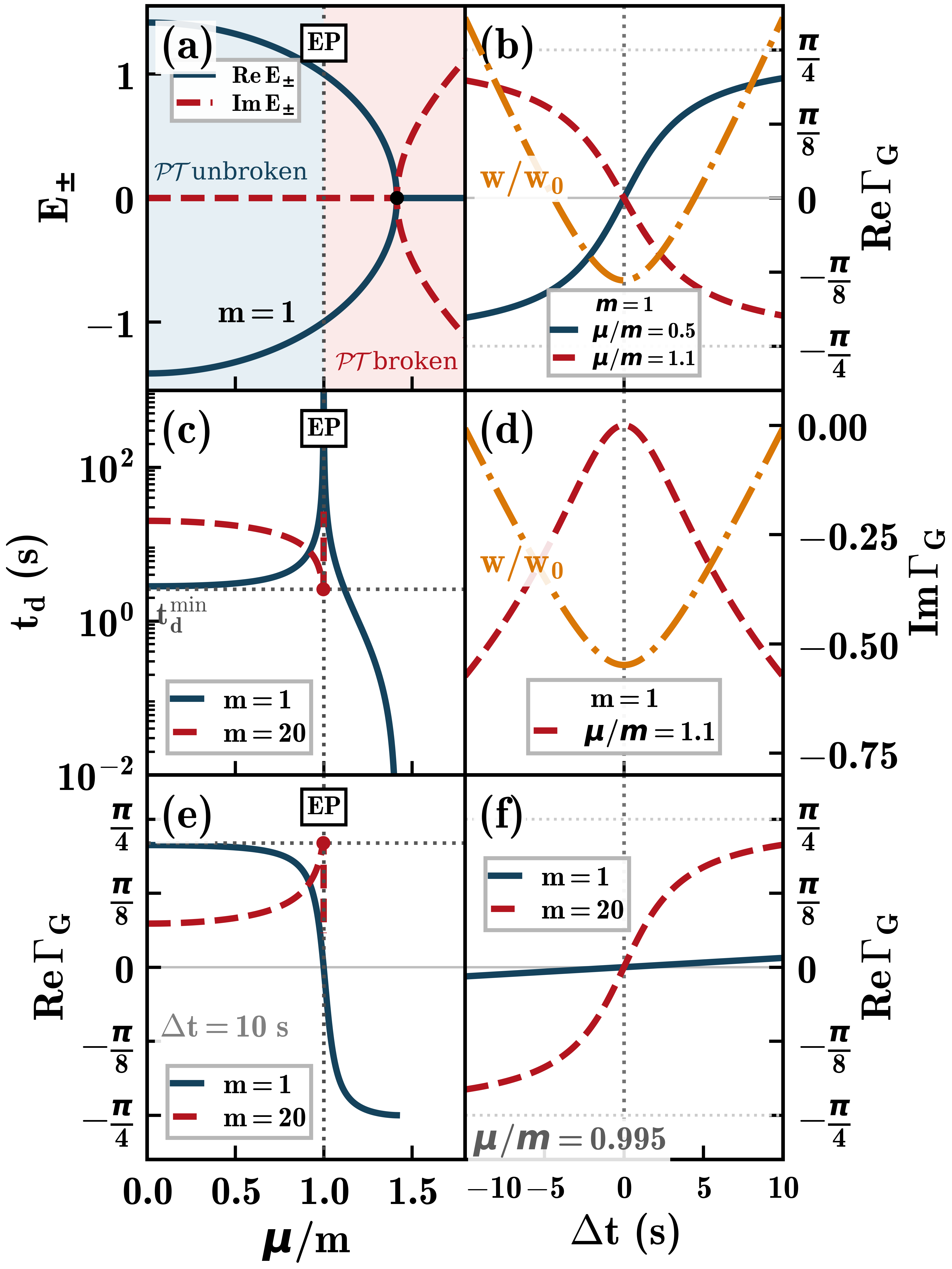}
\caption{Gouy phase across the $\mathcal{PT}$-symmetry-breaking transition; the left column shows the dependence on $\mu/m$, while the right column shows the dependence on the propagation time $\Delta t=t-t_0$, where $t_0$ denotes the time at which the wavepacket reaches its minimum width, with each column sharing a single horizontal axis. (a) Energy spectrum for $m=p_0=w_0=1$: real in the unbroken regime and above the exceptional point EP up to approximately $\mu/m=1.4$, becoming complex beyond this critical value. (b) Real part of the Gouy phase and packet width $w/w_0$ in the unbroken regime ($\mu/m=0.5$), and Gouy phase in the broken regime ($\mu/m=1.1$), for a light mass $m=1$.
(c) Spreading time $t_d$ for a light $m=1$ and a heavy mass $m=20$ packet; on the unbroken regime the minimum spreading time is $t_d^{\min}=\tfrac{3\sqrt{3}}{2}w_0^{2}p_0$, which is mass-independent. (d) Imaginary part of the Gouy phase and packet width in the broken regime ($\mu/m=1.1$). (e) Real part of the Gouy phase evaluated at $\Delta t=10$~s for small $m=1$ and large mass $m=20$. (f) Real part of the Gouy phase near the exceptional point (EP), at $\mu/m=0.995$, for small and large masses.}
\label{fig:gouypt}
\end{figure}

Before detailing the geometric interpretation, it is crucial to specify that in our model $\mathcal{PT}$-unbroken and $\mathcal{PT}$-broken refer to the spectral and eigenstate realization of the $\mathcal{PT}$-invariant
Hamiltonian. Thus, $\mathcal{PT}$ symmetry corresponds to the unbroken regime where the non-Hermitian parameter $\mu$ is smaller than the mass scale $m$ ($|\mu| < |m|$), preserving a purely real spectrum. As illustrated in Fig.~\ref{fig:gouypt}, the transition across the exceptional point is directly reflected in the Gouy phase: its real part undergoes a sign reversal [Fig.~\ref{fig:gouypt}(b)], accompanied by the emergence of a complex part [Fig.~\ref{fig:gouypt}(d)]. The wavepacket width $w(t)$ remains real in both regimes. As seen in Fig.~\ref{fig:gouypt}(c), in the $\mathcal{PT}$-unbroken regime the spreading time attains a minimum value $t_d^{\min}$ which is mass-independent. In the limit of small mass, the spreading time $t_d$ increases from its minimum value and diverges as the EP is approached. Beyond the EP, in the broken regime, $t_d$ decreases until reaching a critical value, which depends on the specific parameters of the system and, for the parameters considered here, occurs around $\mu/m \simeq 1.4$, beyond which becomes complex. In contrast, in the large-mass limit, $t_d$ decreases in the unbroken regime toward its minimum value at the EP and becomes complex immediately upon entering the broken regime. The behavior of the spreading time $t_d$ shown in Fig.~\ref{fig:gouypt}(c) directly explains the evolution of the real part of the Gouy phase displayed in Fig. ~\ref{fig:gouypt}(e) and Fig.~\ref{fig:gouypt}(f). In the small-mass limit, the Gouy phase decreases throughout the unbroken regime from a value close to $\pi/4$, corresponding to the phase at $\Delta t=10\;\mathrm{s}$ in Fig.~\ref{fig:gouypt}(b), and vanishes at the EP. Upon entering the broken regime, the phase changes sign and rapidly approaches $-\pi/4$, as $t_d$ continues to decrease until it eventually becomes complex. In contrast, in the large-mass limit, the Gouy phase increases throughout the unbroken regime, approaching $\pi/4$ at the EP. Beyond the EP, no real Gouy-phase value is displayed, consistently with $t_d$ becoming complex in the broken regime. In Fig.~\ref{fig:gouypt}(f), we show the Gouy phase as a function of the evolution time $\Delta t$ near the exceptional point (EP), considering both the small- and large-mass limits. A striking contrast emerges between these two regimes: while the Gouy phase exhibits only a very small variation for small masses, its variation becomes pronounced in the large-mass limit. This counterintuitive behavior represents an exotic feature of the EP, where the usual expectation associated with the mass dependence is effectively reversed. In particular, the enhanced Gouy-phase variation for large masses suggests a more pronounced quantum behavior, whereas the small-mass limit displays an effectively more classical response, with an almost negligible phase variation.

The contrasting behavior of $t_d$ follows directly from the carrier energy $E_0=\sqrt{p_0^2+m^2-\mu^2}$, since $\beta_2\propto E_0^{-3}$ and therefore $t_d\propto E_0^3$. In the small-mass limit, the momentum contribution $p_0^2$ helps keep $E_0$ real beyond the EP, as shown in Fig.~\ref{fig:gouypt}(a), allowing $t_d$ to remain real in part of the broken regime until a critical value of $\mu$ is reached. In contrast, for large masses, $p_0^2$ is comparatively small relative to $m^2$ and $\mu^2$, so that $E_0$ becomes imaginary almost immediately beyond the EP, making $t_d$ complex and explaining the abrupt termination of the corresponding curve in Fig.~\ref{fig:gouypt}(c).

\textit{Perspectives for Experimental Detection of the Gouy Phase} -  The complex nature of the Gouy phase in the broken regime provides an operational route to measure both its real and imaginary components. Non-Hermitian photonics provides a direct experimental setting for effective non-Hermitian Hamiltonians, because
paraxial light propagation in waveguide or resonator platforms is governed by a Schr\"odinger-type equation $i\partial_z \bm{\psi}(z) = \mathcal{H}_{\rm opt}\,\bm{\psi}(z)$, so our time-evolution results can be carried over to optics under the identification $t\leftrightarrow z$ and $H\leftrightarrow \mathcal{H}_{\rm opt}$ in the parameter regime where the evolution is quasi-Hermitian. In this regime, the imaginary part of the Gouy phase, $\Im\Gamma_G(z)$, encodes the non-unitary attenuation of the wavepacket [See Eq.~\eqref{eq:peak_amplitude}]. To extract this non-Hermitian contribution, one must isolate it from the spatial spreading of the wavepacket. As schematically illustrated in Fig.~\ref{fig:setup}, this can be experimentally realized by injecting a wavepacket into a planar lossy waveguide and capturing its intensity profile with a detection system to extract its attenuation. The magnitude of the peak amplitude $\vert{}A(z)\vert{}$ of the wavepacket evolves as a result of the diffractive broadening along the $x$ direction (which can be directly assessed via the wavepacket width $w(z)$) and the non-Hermitian exponential factor:

\begin{figure}[htb]
\centering\includegraphics[scale=0.47]{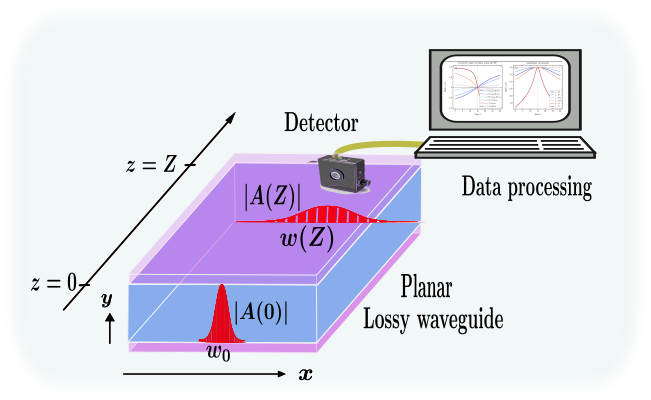}
\caption{Proposed experimental setup for the detection of the non-Hermitian Gouy phase. An initial optical wavepacket with known amplitude $\vert{}A(0)\vert{}$ and width $w_0$ is injected into a planar lossy waveguide. After propagating until $z=Z$, the beam's intensity is captured by a detection system, yielding the final amplitude $\vert{}A(Z)\vert{}$ and wavepacket width $w(Z)$. These observables enable the extraction of the complex phase components discussed above.}\label{fig:setup}\end{figure}

\begin{equation}|A(z)| = |A(0)| \sqrt{\frac{w_0}{w(z)}} \exp[\Im\Gamma_G(z)].\label{eq:peak_amplitude}\end{equation} By measuring the relative reduction in the amplitude magnitude over $z$ direction and concurrently tracking the width evolution $w(z)/w_0$, the imaginary component of the Gouy phase can be determined from:\begin{equation}\Im\Gamma_G(z) = \ln\left[ \frac{|A(z)|}{|A(0)|} \right] + \frac{1}{2} \ln\left[ \frac{w(z)}{w_0} \right].\end{equation} Once $\Im\Gamma_G(z)$ is mapped, the real part of the Gouy phase, $\Re\Gamma_G(z)$, can be determined without the need for delicate interferometric setups that demand stabilization cost. By invoking the Kramers-Kronig relations~\cite{Yariv1989} and evaluating the Cauchy principal value integral of the imaginary phase over the parameter $z^{\prime}$, the real Gouy phase component is given by:\begin{equation}\Re\Gamma_G(z) = -\frac{1}{\pi} \mathcal{P} \int_{-\infty}^{\infty} \frac{\Im\Gamma_G(z^{\prime})}{z^{\prime} - z} dz^{\prime} = \frac{1}{2}\arctan\left(\frac{z}{z_d}\right),\end{equation} where $z_d=\pi w_0^{2}/\lambda$ is the Rayleigh length, with $\lambda$ denoting the wavelength of light. This mapping demonstrates that integrating properly normalized experimental attenuation data recovers the real phase shift.

\textit{Conclusions and outlook} - 
Summarizing, we have shown that the Gouy phase provides a sensitive probe of $\mathcal{PT}$-symmetry breaking in non-Hermitian Dirac dynamics. While it remains real in the unbroken regime, beyond the exceptional point, it changes sign and acquires an imaginary component. At the exceptional point, its variation vanishes for small masses but becomes maximal for large masses, revealing a counterintuitive crossover from effectively classical to increasingly wave-like quantum behavior. We also propose an optical scheme to access the Gouy phase through attenuation in a lossy waveguide. These results establish the non-Hermitian Gouy phase as a geometric probe of exceptional-point physics with potential applications in photonic systems with gain and loss. Our results also suggest several directions for future work. It would be interesting to establish a closer connection with the Lindbladian description of open quantum systems for matter waves, and to explore a non-Hermitian generalization of the Bargmann invariant associated with the Gouy phase. Further insight may also come from studying the relation between the non-Hermitian Gouy phase and evanescent waves.
\vspace{0.3cm}

\begin{acknowledgments}
D.J.C.M.S. acknowledges CAPES (Brazil) for scholarship support. I.G.P. acknowledges Grant No. 306528/2023-1 from CNPq. L.S.M. acknowledges Grant No. 3305943/2026-0 from CNPq. G. P. acknowledges support from national funds by FCT - Fundação para a Ciência e Tecnologia, I.P. in the framework of the project UID/04564/2025, with DOI identifier 10.54499/UID/04564/2025.
\end{acknowledgments}

\bibliography{references.bib}

\end{document}